\documentclass[11pt,a4paper,fleqn]{article}
\usepackage{graphicx}
\usepackage{psfrag}
\usepackage{amsmath}

\newcommand{\figpath}{figures}

\newcommand{\dt}{\delta t}
\newcommand{\DT}{\Delta T}

\newcommand{\eff}{\mbox{\scriptsize eff}}

\newcommand{\sigmaEff}{\sigma_{\eff}}

\newcommand{\wInfty}{w_\infty}
\newcommand{\vInfty}{v_\infty}

\newcommand{\sigmaInfty}{\sigma_\infty}

\newcommand{\E}[1]{E\left[~ #1 ~\right]}
\newcommand{\condExpt}[1]{ E\left[#1 ~|~ \Omega(t)\right]}

\begin{document}
\thispagestyle{empty}

\begin{center}
{\LARGE\bf Volatility forecasts and the at-the-money implied volatility: \\
	a multi-components ARCH approach \\[0.8ex] and its relation with market models}

\vspace*{3ex}
\parbox{0.35\textwidth}{\renewcommand{\baselinestretch}{1.0}\normalsize
{\bf Gilles Zumbach} \\[2ex]
RiskMetrics Group \\
Av. des Morgines 12 \\
1213 Petit-Lancy\\
Geneva, Switzerland\\[1.5ex]
gilles.zumbach@riskmetrics.com}

\vspace*{2ex}
First version: December 17, 2007\\
This version: \today

\vspace*{3ex}

{\bf Abstract}
\end{center}
For a given time horizon $\DT$, this article explores the relationship between the realized volatility (the volatility that will occur between $t$ and $t+\DT$), the implied volatility (corresponding to at-the-money option with expiry at $t+\DT$), and several forecasts for the volatility build from multi-scales linear ARCH processes. 
The forecasts are derived from the process equations, and the parameters set {\it a priori}. 
An empirical analysis across multiple time horizons $\DT$ shows that a forecast provided by an I-GARCH(1) process (1 time scale) does not capture correctly the dynamic of the realized volatility. 
An I-GARCH(2) process (2 time scales, similar to GARCH(1,1)) is better, while a long memory LM-ARCH process (multiple time scales) replicates correctly the dynamic of the realized volatility and delivers consistently good forecast for the implied volatility. 
The relationship between market models for the forward variance and the volatility forecasts provided by ARCH processes is investigated. 
The structure of the forecast equations is identical, but with different coefficients. 
Yet the process equations for the variance are very different (postulated for a market model, induced by the process equations for an ARCH model), and not of any usual diffusive type when derived from ARCH.

\vspace{4ex} 
\large{RiskMetrics Group}\\
\large{One Chase Manhattan Plaza} \hfill\\
\large{44th Floor} \hfill\\ 
\large{New York, NY 10005 \hfill \texttt{\textbf{www.riskmetrics.com}}}
\newpage

\section{Introduction}
The intuition behind volatility is to measure price fluctuations, or equivalently the typical magnitude for the price changes.
Yet, beyond the first intuition, volatility is a fairly complex concept, for various reasons.  First,  turning this intuition into formulas and numbers is partly arbitrary, and many meaningful and useful definitions of volatilities can be given. 
Second, the volatility is not directly ``observed'' or traded, but rather computed from time series (although this situation is changing indirectly through the ever increasing and sophisticated option market, the volatility indexes and the options on volatility).
For trading strategies, options and risk evaluations, the valuable quantity is the realized volatility, namely the volatility that will occur between the current time $t$ and some time in the future $t+\DT$. As this quantity is not available at time $t$, a forecast needs to be constructed. Clearly, a better forecast of the realized volatility allows to better price options, to make profit on volatility based trades,  and to manage better risks in a portfolio. 

At a time $t$, a forecast for the realized volatility can be constructed from the (underlying) price time series. In this paper, multiscales ARCH processes are used.  On the other hand, a liquid option market allows to compute the implied volatility, corresponding to the ``market'' forecast for the realized volatility. On the theoretical side, an ``instantaneous'', or effective, volatility $\sigmaEff$ is needed to define processes, and the forward variance. Therefore, at a given time $t$, we have mainly one theoretical instantaneous volatility and three notions of ``observable'' volatility (forecasted, implied and realized). This paper studies the empirical relationship between these three time series, as a function of the forecast horizon $\DT$. There exist already an abundant literature on this topic, and \cite{Poon.2003} published a book summarizing nicely the available publications ($\sim$100 articles on volatility forecast alone!).

The main line of this work is to model the underlying time series by multi-components ARCH processes, and to {\em derive} a volatility forecast. 
This forecast, based only on the underlying, should be close to the implied volatility for the at-the-money (ATM) option. 
In particular when option data are poor, lacking or not available, such approach allows to obtain a good approximation for the  ATM implied volatility. 
For trading and risk management, the correct pricing of options is clearly an issue, and to have a fall-back solution for the implied volatility surface using a minimal modeling of the underlying is a clear advantage. 
This article does not address the issue of the full surface, but only the implied volatility for the ATM options, called the {\em backbone}.

A vast literature on implied volatility and its dynamic already exists. In this article, we will review some recent developments on market models for the forward variance. These models focus on the volatility as a process, and many process equations can be set that are compatible with a martingale condition for the volatility. On the other side, the volatility forecast as induced by a multi-components ARCH process leads also to process equations for the volatility only. These two approaches leading to process for the volatility are contrasted, showing the formal similarity in the structure of the forecasts, but the very sharp difference in the processes for the volatility. If the price time series behave according to some ARCH process, then the implication for volatility modeling is far reaching as the usual structure based on Wiener process cannot be used.

This paper is organized as follow. The required definitions for the volatilities and forward variance are given in the next section.
The various multi-components ARCH processes are introduced in sec.~\ref{sec:ARCH}, and the induced volatility forecasts  and processes given in sec~\ref{sec:forwardVarianceAndARCH} and \ref{sec:ARCHinducedVolProcess}.  
The market models and the associated volatility dynamics are presented in sec.~\ref{sec:varianceMarketModel}. 
The relationship between market models, options and the ARCH forecasts are discussed in section~\ref{sec:marketModelsAndOption}.
Section~\ref{sec:empiricalComparison} presents an empirical investigation of the relationship between the forecasted, implied and realized volatilities, before the conclusion.

\section{Definitions and setup of the problem}
\subsection{General}
We assume to be at time $t$, with the corresponding information set $\Omega(t)$.
The time increment for the processes and the granularity of the data is denoted by $\dt$, and is 1 day in the present work.
We assume that there exists an instantaneous volatilities denoted by $\sigmaEff(t)$, 
which corresponds to the annualized expected standard deviation of the price in the next time step $\dt$. This is a usefull quantity for the definitions, but this volatility is essentially unobserved. In a process,  $\sigmaEff$ gives the magnitude of the returns.

\subsection{Realized volatility}
The {\em realized volatility} corresponds to the annualized standard deviation of the returns in the interval between $t$ and $t+\DT$
\begin{equation}
  \sigma^2(t, t+\DT) = \frac{\text{1 year}}{n\;\dt} \sum_{t < t' \leq t+\DT}  r^2(t') 
\end{equation}
where $r(t)$ are the (unannualized) returns measured over the time interval $\dt$, and the ratio 1 year/$\dt$ annualized the volatility. The empirical section is done with daily data and the returns are evaluated over a 1 day interval $\dt$ = 1 day.
If the returns do not overlap in the sum, then $\DT = n\;\dt$.
At the time $t$, the realized volatility cannot be evaluated from the information set $\Omega(t)$. The realized volatility is {\em the} usefull quantity we would like to forecast and to relate to the implied volatility.


\subsection{ Forward variance }
In a continuum time formulation, the {\em expected cumulative variance} is defined by
\begin{equation}
   V(t, t+\DT) = \int_t^{t+\DT} dt' ~\condExpt{\sigmaEff^2(t')} 
\end{equation}
and the {\em forward variance} by
\begin{equation}
   v(t, t+\DT) = \frac{\partial V(t, t+\DT) }{\partial\DT} =  \condExpt{\sigmaEff^2(t+\DT)}.
\end{equation}
The cumulative variance is an extensive quantities as it is proportional to $\DT$.
For empirical investigation, it is simpler to work with an intensive quantity as this remove a trivial dependency on the time horizon. For this reason, the cumulative variance is used only in the theoretical part (hence also the continuum definition with an integral), whereas the forecasted volatility is used in the empirical part.

The variance enters into the variable leg of a variance swap, and as such, it is tradable.
Related tradable instruments are the volatility indexes like the VIX
(but the relation is indirect as the index is defined through implied volatility of a basket of options).
Because volatility is becoming tradable, the forward variance should be a martingale
\begin{equation}
  \condExpt{v(t', T)} = v(t, T).  \label{eq:martingale}
\end{equation}
For the volatility, this condition is quite weak as it
follows also from the chain rule for conditional expectation
\begin{equation}
  \condExpt{ \E{\sigmaEff^2(T) ~|~ \Omega(t')} } = \condExpt{\sigmaEff^2(T)}  \qquad\text{for } t <  t' < T
\end{equation}
and from the definition of the forward variance as a conditional expectation.
Therefore, any forecast build as a conditional expectation produces a martingale for the forward variance.

At this level, there is a formal analogy with interest rates, with the (zero coupon) interest rate and forward rate being analogous to the cumulative variance and forward variance. 
Therefore, some ideas and equations can be borrowed from the IR field.
For example, on the modeling side, one can write process for the cumulative variance or for the forward variance, 
the later being more convenient as the martingale condition gives simpler constraints on the possible equations.
In this paper, the ARCH path is followed using a multi-scale process for the underlying. The forward variance is computed as an expectation, and therefore the martingale property follows. In section \ref{sec:varianceMarketModel}, this ARCH approach is contrasted with a direct model for the forward volatility, where the martingale condition has to be explicitely enforced.

\subsection{The forecasted volatility}
The {\em forecasted volatility} is defined by
\begin{equation}
   \widetilde{\sigma}^2(t, t+\DT) = \frac{1}{n} \sum_{t < t' \leq t+\DT} \condExpt{\sigmaEff^2(t')} 
\end{equation}
Up to a normalization and the transformation of the integral into a discrete sum, this definition is similar to the expected cumulative variance.

\subsection{The implied volatility}
As usual, the implied volatility is defined as the volatility to insert into the Black-Sholes equation so as to recover the market price for the option.
The implied volatility  $\sigma_{BS}(m, \DT)$ is a function 
of the moneyness $m$ and of the time to maturity $\DT$.
The moneyness can be defined is various ways, 
with most definitions similar to $ m \simeq \ln\left(F/K\right)$, and with $F$ the forward rate $F = S e^{r\;\DT}$.
The (forward) at-the-money option corresponds to $m=0$.
The {\em backbone} is the implied volatility at the money $\sigma_{BS}(\DT) = \sigma_{BS}(m = 0, \DT)$, as a function of the time to maturity $\DT$.
For a given time to maturity $\DT$, the implied volatility as function of moneyness is called the smile.

Intuitively, the implied volatility surface can loosely be decomposed in backbone $\times$ smile.
The rationale for this decomposition is that the two directions depend on different option features.
The backbone is related to the expected volatility until the option expiry
\begin{equation}
   \widetilde{\sigma}(t, t+\DT) =  \sigma_{BS}(m = 0, \DT)(t)  \label{eq:backboneVsCumulativeVariance}
\end{equation}
In the Black-Sholes formula,  the volatility appears only through the combination $\DT\;\sigma^2$, corresponding to the cumulative expected variance.
In the other direction, the smile is the fudge factor to remedy the incomplete modeling of the underlying by a Gaussian random walk. The Black-Sholes model has the key advantage to be solvable, but does not include many stylized facts like heteroscedasticity, fat-tails, or leverage effect. These shortcomings translate into various ``features'' of the smile.

In principle, the equation \ref{eq:backboneVsCumulativeVariance} should be checked using empirical data.
Yet this comparison raises a number of issues, on both sides of the equation.
On the left hand side, the variance forecast should be computed using some equations and the time series for the underlying. The forecasting scheme, with its estimated parameters, is subject to errors.
On the right had side, the option market has its own idiosyncracies, for example related to demand and supply.
Such effect can be clearly observed by computing the implied volatility corresponding to the option bid or ask prices.
These points are discussed in more details in sec.~\ref{sec:empiricalComparison}.
Therefore, the equation \ref{eq:backboneVsCumulativeVariance} should be taken only as a first order approximation.

\section{Multi-components  ARCH processes}
\label{sec:ARCH}
\subsection{The general setup}
The basic idea of a multi-components ARCH process is to measure historical volatilities using exponential moving average on a set of time horizons, and to compute the effective volatility for the next time step as a convex combination of the historical volatilities.
A first process along similar line was introduced in \cite{Dacorogna.1998}, and this family of processes was throughly developed and  explored in \cite{ZumbachLynch, LynchZumbach, Zumbach.LongMemory}. 
A particular simple process with long memory is used to build the RM2006 risk methodology \cite{Zumbach.RM2006_fullReport}, with the salient feature to be very parsimonious. 
One of the key advantage of these multi-components processes is that forecast for the variance can be computed analytically. 
We will use this property to explore their relations with the option implied volatility.

In order to build the process, 
the historical volatilities are measured by exponential moving averages (EMA) at time scales $\tau_k$
\begin{equation}
    \sigma_k^2(t) = \mu_k ~\sigma_k^2(t-\dt) + (1 - \mu_k) ~r^2(t)  \hspace{3em}k = 1, \cdots, n
    \label{eq:sigma_k}
\end{equation}
and with decay coefficients $\mu_k = \exp( -\dt/\tau_k)$. The process time increment is $\dt$, and $\dt$ = 1 day in this work. 
Let us emphasize that the $\sigma_k$ are computed from historical data, and there is no hidden stochastic processes like in a stochastic volatility model.

The ``effective'' variance $\sigmaEff^2$ is a convex combination of the 
$\sigma_k^2$ and of the mean variance $\sigmaInfty^2$
\begin{eqnarray}
    \sigmaEff^2(t) & = &  \sum_{k = 1}^n w_k ~\sigma_k^2(t)  + \wInfty ~\sigmaInfty^2
   ~=~ \sigmaInfty^2 + \sum_{k = 1}^n w_k ~\left(\sigma_k^2(t) - \sigmaInfty^2 \right) \\
    1 & = &  \sum_{k = 1}^n w_k + \wInfty   \nonumber
\end{eqnarray}
Finally, the price follow a random walk with volatility $\sigmaEff$
\begin{equation}
	r(t+\dt) = \sigmaEff(t) ~\epsilon(t+\dt).
\end{equation}
Depending on the number of components $n$, the time horizons $\tau_k$ and weights $w_k$, a number of interesting processes can be build. The processes we are using to compare with implied volatility are given in the next subsections.

On general ground, we make the distinction between affine processes for which the mean volatility is fixed by $\sigmaInfty$ and $\wInfty > 0$, and the linear process for which $\wInfty = 0$. 
The linear and affine terms qualify the equations for the variance (i.e. in $\sigma^2$). 
The linear processes are very interesting for forecasting volatility as they have no mean volatility parameter $\sigmaInfty$ which is clearly time series dependent. However, their asymptotic properties are singular, and affine processes should be used in Monte Carlo simulations. This subtle difference between both classes of processes is discussed in details in \cite{Zumbach.LongMemory}. As this paper deal with volatility forecasts, only the linear processes are used. 

\subsection{I-GARCH(1)}
The I-GARCH(1) model corresponds to a 1-component linear process
\begin{eqnarray*}
    \sigma^2(t) & = & \mu ~\sigma^2(t-\dt) + (1 - \mu) \,r^2(t)  \\[1ex]
   \sigmaEff^2(t) & = & \sigma^2(t).
\end{eqnarray*}
It has one parameter $\tau$ (or equivalently $\mu$).
This process is equivalent to the integrated GARCH(1,1) process \cite{EngleBollerslev.1986},
and with a given value for $\mu$ is equivalent to the standard RiskMetrics methodology.
Its advantage is to be the most simple, but it does not capture mean revertion for the forecast (i.e. that forecasts for increasing horizons should converge to a (mean) long term volatility).

For the empirical evaluation, the characteristic time has been fixed {\it a priori} to $\tau$ = 16 business days, corresponding to $\mu \simeq 0.94$.

\subsection{I-GARCH(2) and  GARCH(1,1)}
The I-GARCH(2) process corresponds to a 2-components linear model
\begin{eqnarray}
    \sigma_1^2(t) & = & \mu_1 ~\sigma_1^2(t-\dt) + (1 - \mu_1) \,r^2(t)  \nonumber\\[1ex]
    \sigma_2^2(t) & = & \mu_2 ~\sigma_2^2(t-\dt) + (1 - \mu_2) \,r^2(t)  \\[1ex]
   \sigmaEff^2(t) & = & w_1\sigma_1^2(t) + w_2\sigma_2^2(t)  \nonumber
\end{eqnarray}
It has three parameters $\tau_1$, $\tau_2$ and $w_1$.
Even if this process is linear, it has mean reversion for time scale up to $\tau_2$, with $\sigma_2(t)$ playing the role of the mean volatility. 

The GARCH(1,1) process \cite{EngleBollerslev.1986} corresponds to the 1-component affine model
\begin{eqnarray}
    \sigma_1^2(t) & = & \mu_1 ~\sigma_1^2(t-\dt) + (1 - \mu_1) \,r^2(t)  \\[1ex]
   \sigmaEff^2(t) & = &  (1 - \wInfty)\,\sigma_1^2(t) + \wInfty \sigmaInfty^2  \nonumber
\end{eqnarray}
It has three parameters $\tau_1$, $\wInfty$ and $\sigmaInfty$.
In this form, the analogy between the I-GARCH(2) and GARCH(1,1) processes is clear, with the long term volatility $\sigma_2$ playing a similar role as the mean volatility $\sigmaInfty$. 

Given a process, the parameters need to be estimated on a time series. GARCH(1,1) is more problematic with that respect because $\sigmaInfty$ is clearly time series dependent. A good procedure is to estimate the parameters on a moving historical sample, say in a window between $t-\DT'$ and $t$ for a fixed span $\DT'$. With this setup, the mean variance $\sigmaInfty^2$ is essentially the sample variance $\sum r^2$ computed on the estimating window.  This is a rectangular moving average, similar to an EMA but for the weights given to the past. This argument shows that I-GARCH(2) and (a continuously re-estimated on a moving window) GARCH(1,1) behaves similarly. A detailled analysis of both processes in \cite{Zumbach.LongMemory} show that they have similar forecasting power, with an advantage to I-GARCH(2).

In this work, we use the I-GARCH(2) process with two parameter sets fixed {\it a priori} to some reasonable values.
The first set is $\tau_1$ = 4 business days, $\tau_2$ = 512 business days, $w_1$ = 0.843 and $w_2$ = 0.157. 
The second set is $\tau_1$ = 16 business days, $\tau_2$ = 512 business days, $w_1$ = 0.804 and $w_2$ = 0.196.
The values for the weights are obtained according to the long memory ARCH process, but with only two given $\tau$ components.

\subsection{Long Memory ARCH}
The idea for a long memory process is to use a multi-components ARCH model with a large number of components but simple analytical form for the characteristic time $\tau_k$ and the weights $w_k$.
For the long memory ARCH process, the characteristic times $\tau_k$ increase as a geometric series
\begin{equation}
    \tau_k   = \tau_1 ~\rho^{k-1}  \hspace{2em} k = 1,\cdots, n  
\end{equation}
while the weights decay logarithmically
\begin{eqnarray}
     w_k  & = & \frac{1}{C} ~\left( 1 - \ln(\tau_k)/\ln(\tau_0)\right) \\
     C  & = & \sum_k \left( 1 - \ln(\tau_k)/\ln(\tau_0)\right).  \nonumber
\end{eqnarray}
This choice produces lagged correlations for the volatility that decays logarithmically, as observed in the empirical data \cite{Zumbach.RM2006_fullReport}.
The parameters are taken as for the RM2006 methodology \cite{Zumbach.RM2006_fullReport}, namely $\tau_1$ = 4 business days, $\tau_n$ = 512 business days, $\rho = \sqrt{2}$ and the logarithmic decay factor $\tau_0$ = 1560 days = 6 years .

\section{Forward variance and multi-components  ARCH processes}
\label{sec:forwardVarianceAndARCH}
For multiscales ARCH processes (I-GARCH, GARCH(1,1), long-memory ARCH, etc ...),
the forward variance can be computed analytically \cite{Zumbach.LongMemory, Zumbach.RM2006_fullReport}.
The idea is to compute the conditional expectation of the process equations, 
from which iterative relations can be deduced. Then, some algebra and matrix computations allow to get the following form for the forward variance
\begin{equation}
    v(t, t+\DT) = \condExpt{\sigmaEff^2(t+\DT)}
	 =  \sigmaInfty^2 + \sum_{k= 1}^n w_k(\DT) \left( \sigma_k^2(t) -  \sigmaInfty^2 \right)
   \label{eq:ARCHForecast}
\end{equation}
The weight $w_k(\DT)$ can be computed by a recursion formula depending on the decay coefficients $\mu_k$ and with initial values given by $w_k = w_k(1)$. The equation for the forecast of the realized volatility has the same form but the weights $w_k(\DT)$ are different.

Let us emphasize that this can be done for all processes in this class (linear and affine).
Moreover, the $\sigma_k^2(t)$ are computed from the underlying time series, namely there is no hidden stochastic volatility to estimate. This makes volatility forecasts particularly easy in this framework.

\begin{figure}
 \begin{center}
  \psfrag{time}{Time interval $\DT$  ~~~[day]}
  \psfrag{weights}{weights $w_k(\DT)$}
  \includegraphics[clip, width = 0.75\textwidth]{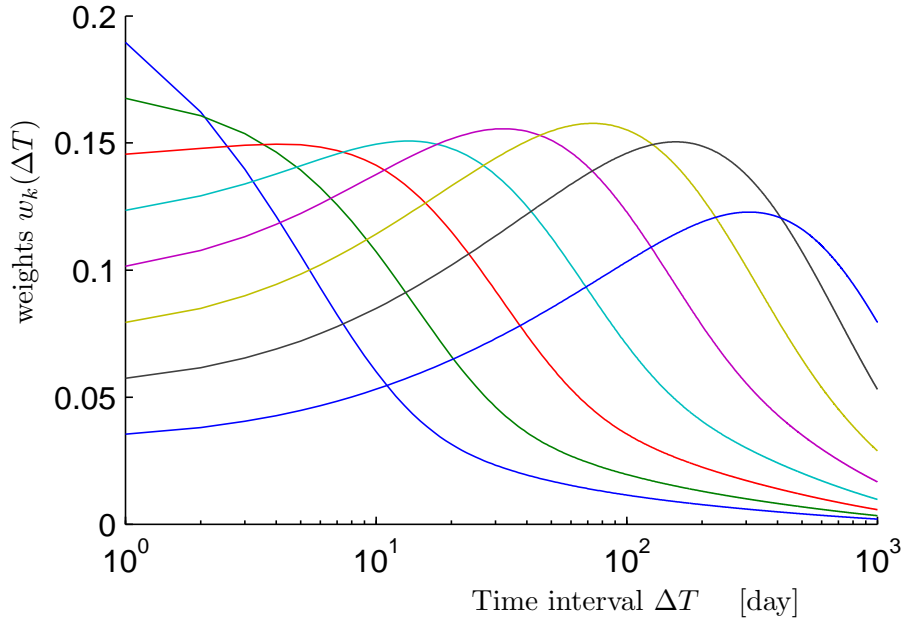}
  \end{center}
 \caption{The weights $w_k(\DT)$ as function of the forecst horizon $\DT$ for a long memory process with $\wInfty = 0.1$ and  $\tau_k = 2, 4, 8, 16,\cdots, 256$ days .
	The weights with  increasing time horizon $\tau_k$ have decreasing initial values and the maximum values going from left to right.}
    \label{fig:forecastWeights}
\end{figure}
\begin{figure}
 \begin{center}
  \psfrag{time}{Time interval $\DT$  ~~~[day]}
  \includegraphics[clip, width = 0.75\textwidth]{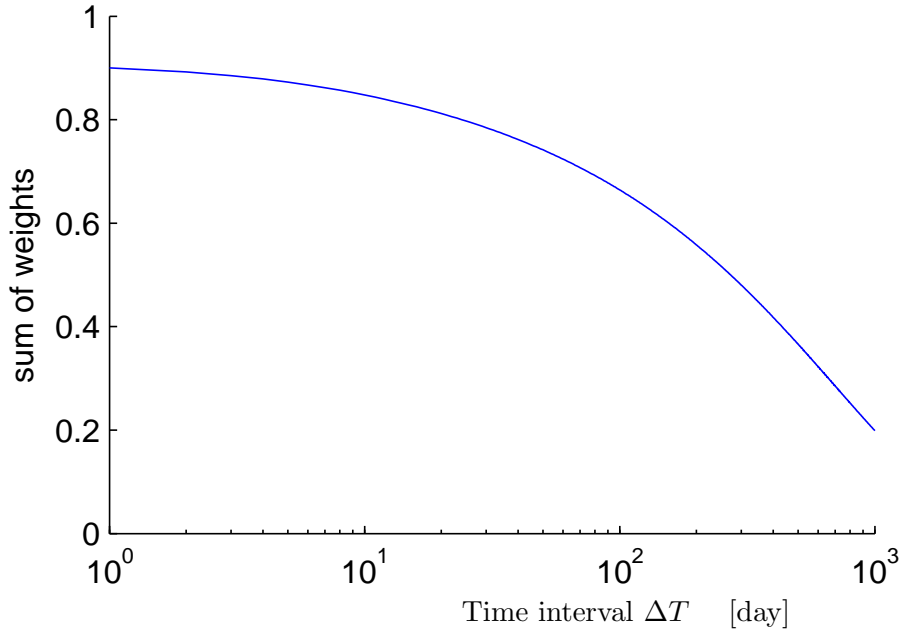}
 \end{center}
  \caption{The sum of the weights $\sum_k w_k(\DT) = 1 - \wInfty$, for the same parameters as above.}
  \label{fig:sumWeights}
\end{figure}

For a multi-component ARCH process, the intuition for the forecast can be understood from a graph of the weights $w_k(\DT)$ as function of the forecast horizon $\DT$ as given in Fig.~\ref{fig:forecastWeights}.
For short forecast horizon, the volatilities with the shorter time horizons dominate.
As the forecast horizon get larger, the weights of the short term volatilities decay while the weights of the longer time horizons get larger. The weight for a particular horizon $\tau_k$ peaks at a forecast horizon similar to $\tau_k$, for example the Burgundy curve corresponds to $\tau = 32$ days and its maximum is around  a similar value. 
The figure~\ref{fig:sumWeights} shows the sum of the volatility coefficients $\sum_k w_k = 1 - \wInfty$. This shows the increasing weight of the mean volatility as the forecast horizon get longer. 
Notice that this behavior corresponds to our general intuition about forecasts, namely short term forecasts depend mainly on the recent past while long term forecasts need to use more informations from the distant past. 
The nice feature of the multi-components ARCH process is that the forecast weights are derived from the process equations, and that they have a similar content compared to the process equations (linear or affine, one or multiple time scales).

\section{The induced volatility process}
\label{sec:ARCHinducedVolProcess}
The multi-components ARCH processes are stochastic processes for the return, in which the volatilities are convenient intermediate quantities. It is important to realize that the volatilities $\sigma_k$ and $\sigmaEff$ are useful and intuitive in formulating a model, but they can be completely eliminated from the equations. An important advantage of this class of process is that the forward variance $v(t, t+\DT)$ can be computed analytically. Going in the opposite direction, we want to eliminate the return, namely to derive the equivalent process equations for the dynamic of the forward variance induced by a multi-component ARCH process. This will allow us to make contact with some models for the forward variance that are available in the literature and presented in the next section.

The eq.~\ref{eq:sigma_k} for $\sigma_k$ can be rewritten as
\begin{eqnarray}
  d\sigma_k^2(t) & = & \sigma_k^2(t) - \sigma_k^2(t-\dt) \\
	& = & (1 -\mu_k) \left\{-\sigma_k^2(t-\dt) +  ~\epsilon^2(t) ~\sigmaEff^2(t-\dt) \right\} \nonumber\\
	& = & (1 -\mu_k) \left\{\sigmaEff^2(t-\dt) -\sigma_k^2(t-\dt) +  ~(\epsilon^2(t) - 1) ~\sigmaEff^2(t-\dt) \right\} \nonumber
\end{eqnarray}
The equation can be simplified by introducing the annualized variances $v_k = 1y/\dt ~\sigma_k^2$, $v_{\eff} = 1y/\dt ~\sigmaEff^2$ and a new random variable $\chi$ with
\begin{equation}
  \chi = \epsilon^2 - 1 \qquad\qquad\text{such that}\qquad \E{\chi(t)} = 0, ~~\chi(t) > -1.
\end{equation}
Assuming that the time increment $\dt$ is small compared to the time scales $\tau_k$ in the model, the following approximation can be used
\begin{equation}
  1 - \mu_k = \frac{\dt}{\tau_k} + O(\dt^2).
\end{equation}
In the present derivation, this expansion is used only to make contact with the more usual continuous time form, but no term of higher order are neglected. Exact expressions are obtained by replacing $\dt/\tau_k$ by $1 - \mu_k$ in the equations below.

These notations and approximations allows to write the equivalent equations
\begin{subequations}  \label{eq:ARCHvolatility}
\begin{eqnarray}
  dv_k     & = & \frac{\dt}{\tau_k}\left\{v_{\eff} - v_k + \chi \; v_{\eff} \right\}  \label{eq:dv_k}\\[1ex] \label{eq:dvk}
  v_{\eff} & = & \sum_k w_k~v_k + \wInfty v_\infty  \label{eq:v_eff}
\end{eqnarray}
\end{subequations}
The process for the forward variance is given by
\begin{equation}
  dv_{\DT} = \sum_k w_k(\DT) ~dv_k  \label{eq:dv}
\end{equation}
with $dv_\tau(t) = v(t, t+\DT) - v(t-\dt, t-\dt+\DT)$.

The content of Eq.~\ref{eq:dv_k} is the following. The term $\dt\left\{v_{\eff} - v_k\right\}/\tau_k$ gives a mean reversion toward the current effective volatility $v_{\eff}$ at a time scale $\tau_k$. This structure is fairly standard, except for $v_{\eff}$ which is given by a convex combination of all the variances $v_k$. Then, the random term is unusual.
All the variances share the same random factor $\dt \;\chi/\tau_k$, which has a standard deviation of order $\dt$ instead of the usual $\sqrt{\dt}$ appearing in Gaussian model. 

An interesting property of this equation is to enforce positivity for $v_k$ through a somewhat peculiar mechanism.
The equation \ref{eq:dv_k} can be rewritten as
\begin{equation}
    dv_k     = \frac{\dt}{\tau_k}\left\{ - v_k + (\chi + 1) v_{\eff} \right\}
\end{equation}
Because $\chi \geq 1$, the term $(\chi + 1)v_{\eff}$ is never negative, and as  $\dt \;v_k(t-\dt)/\tau_k$ is smaller than $v_k(t-\dt)$, 
this implies that  $v_k(t)$ is always positive (even for a finite $\dt$).
Another difference with the usual random process is that the distribution for $\chi$ is not Gaussian.
In particularly if $\epsilon$ has a fat-tail distribution, as seems required in order to have a data generating process that reproduce the properties of the empirical time series, the distribution for $\chi$ also has fat tails.

The continuum limit of the GARCH(1,1) process was already investigated by \cite{Nelson.1990}. In this limit, GARCH(1,1) is equivalent to a stochastic volatility process where the variance has its own source of randomness. Yet Nelson constructed a different limit as above because he fixes the GARCH parameters $\alpha_0$, $\alpha_1$ and $\beta_1$. The decay coefficient is given by $\alpha_1 + \beta_1 = \mu$ and is therefore fixed. With $\mu = \exp( -\dt/\tau)$, fixing $\mu$ and taking the limit $\dt \rightarrow 0$ is equivalent to  $\tau \rightarrow 0$. Because the characteristic time $\tau$ of the EMA go to zero, the volatility process becomes independent of the return process, and the model converges toward a stochastic volatility model. A more interesting limit is to take $\tau$ fixed and $\dt \rightarrow 0$, as in the computation above.
Notice that the computation is done with a finite time increment $\dt$; the existence of a proper continuum limit $\dt \rightarrow 0$ for a process defined by eq.~\ref{eq:dvk} to \ref{eq:dv} is likely not a simple question.  

Let us emphasize that the derivation of the volatility process as induced by the ARCH structure involves only elementary algebra. Essentially, if the price follows an ARCH process (one or multiple time scales, with or without mean $\sigmaInfty$), then the volatility follows a process according to \ref{eq:ARCHvolatility}. The structure of this process involves a random term of order $\dt$ and therefore it cannot be reduced to a Wiener process. This is a key difference from the processes used in finance that were developed to capture the price diffusion. 

The implications of eq.~\ref{eq:ARCHvolatility} are important as they show a key difference between ARCH and stochastic volatility processes. This has clearly implication for option pricing, but also for risk evaluation. In a risk context, the implied volatility is a risk factor for any portfolio that contains options, and it is likely better to model the dynamic of the implied volatility by a process with a similar structure. 

\section{ Market model for the variance}
\label{sec:varianceMarketModel}
In the literature, the models for the implied volatility are dominated by stochastic volatility processes, essentially assuming that the implied volatility ``has its own life'', independently of the underlying. In this vast literature, a recent direction is to write processes directly for the forward variance. Recent papers in this direction include \cite{Buehler} and \cite{Bergomi}, and a presentation by \cite{Gatheral}. In this direction, we present here simple linear processes for the forward variance, and discuss the relation with a multi-components ARCH in the next section.

The general idea is to write a model for the forward variance
\begin{equation}
    v(t, t+\DT) = G(v_k(t); \DT)
\end{equation}
where $G$ is a given function of the (hidden) random factors $v_k$. In principle, the random factors can appear everywhere in the equation, say for example as a random characteristic time like $\tau_k$. Yet, Buehler has showed that strong constraints exist on the possible random factors, for example forbiding random characteristic time. In this paper, only linear model will be discussed, and therefore the random factor appears as a variance $v_k$.

The dynamic for the random factor $v_k$ are given by processes
\begin{equation}
  dv_k = \mu_k(v) ~dt + \sum_{\alpha=1}^d \sigma_k^\alpha(v) ~dW^\alpha  \hspace{4em}k = 1, \cdots, n.
  \label{eq:dv_k_general}
\end{equation}
The processes have $d$ sources of randomness $dW^\alpha$, and the volatility $\sigma_k^\alpha(v)$ can be any function of the factors. 

As such, the model is essentially unconstraint, but the martingale condition \ref{eq:martingale} for the forward variance still has to be enforced. 
Through standard Ito calculus, the variance curve model together with the martingale condition lead to a constraint
between $G(v; \DT)$, $\mu(v)$ and $\sigma(v)$
\begin{equation}
     \partial_{\DT} G(v; \DT) = \sum_{i= 1}^n \mu_i ~\partial_{v_i} G(v; \DT)
  	+ \sum_{i,j = 1}^n \sum_{\alpha=1}^d \sigma_i^\alpha \sigma_j^\alpha ~~\partial_{v_i, v_j}^2 G(v; \DT)
\end{equation}
A given function $G$ is say to be compatible with a dynamic for the factors if this condition is valid.
The compatibility constraint is fairly weak, and many processes can be written for the forward variance that are martingale.
As already mentionned, we consider only functions $G$ that are linear in the risk factors. Therefore, $\partial_{v_i, v_j}^2 G = 0$, leading to first order differential equations that can be solved by elementary techniques. For this class of models, the condition does not involve the volatility $\sigma_k^\alpha(v)$ of the factor, which therefore can be chosen freely.

\subsection{Example: one factor market model}
The forward variance is parameterized by
\begin{eqnarray}
     G(v_1; \DT) & = & \vInfty + w_1(\DT) (v_1 - \vInfty) \\
	w_1(\DT) & = & w_1 \;e^{-\DT/\tau_1}  \nonumber
\end{eqnarray}
which is compatible with the stochastic volatility dynamic
\begin{equation}
  dv_1 = -(v_1 - \vInfty) ~\frac{dt}{\tau_1} +  \gamma ~v_1^\beta ~dW  \hspace{3em}\text{for}~~\beta \in [1/2, 1].
\end{equation}
The parameter $w_1$ can be chosen freely, and for identification purpose the choice $w_1 = 1$ is often made.
Because $G$ is linear in $v_1$, there is no constraint on $\beta$.
The value $\beta = 1/2$ corresponds to the Heston model, $\beta = 1$ to the log-normal model.
This model is somewhat similar to the GARCH process, with one characteristic time $\tau_1$, a mean volatility $\vInfty$, and the volatility of the volatility (vol-of-vol) $\gamma$.
This model is not rich enough to describe the empirical forward variance dynamic, which involve multiple time scale.

\subsection{Example: two factors market model}
The linear model with two factors
\begin{eqnarray}
     G(v; \DT) & = &  \vInfty + w_1(\DT) ~\left(v_1 - \vInfty\right) +  w_2(\DT) ~\left( v_2 - \vInfty \right)   \label{eq:twoCompStochVol} \nonumber\\
	w_1(\DT) & = & w_1 \;e^{-\DT/\tau_1}  \\
	w_2(\DT) & = & \frac{1}{1 - \tau_1/\tau_2}\left( - w_1\;e^{-\DT/\tau_1} + (w_1 + w_2) \;e^{-\DT/\tau_2} \right) \nonumber
\end{eqnarray}
is compatible with the dynamic
\begin{eqnarray}
  dv_1 & = & -(v_1 - v_2) ~dt/\tau_1 +  \gamma ~v_1^\beta ~dW_1  \\
  dv_2 & = & -(v_2 - \vInfty) ~dt/\tau_2 +  \gamma ~v_2^\beta ~dW_2.  \nonumber
\end{eqnarray}
The parameters $w_1$ and $w_2$ can be chosen freely, 
and for identification purpose the choice $w_1 = 1$ and $w_2 = 0$ is often made.
Notice the similarity of the equation~\ref{eq:twoCompStochVol} with the Nelson-Siegel-Svensson parameterization for the yield curve.

The linear model can be solved explicitely for $n$-components, but the $\DT$ dependency in the coefficients $w_k(\DT)$ becomes increasingly complex. It is therefore not natural in this approach to create the equivalent of a long-memory model with multiple time scales.

\section{Market models and options}
\label{sec:marketModelsAndOption}
Assuming a liquid option market, the implied volatility surface can be extracted, 
and from its backbone, the forward variance $v(t, t+\DT)$ is computed.
At a given time $t$, given a market model $G(v_k(t); \DT)$, the risk factors $v_k(t)$ are estimated by fitting the function $G(\DT)$ on the forward variance curve. It is therefore important for the function $G(\DT)$ to have enough possible shapes to accommodate the various forward variance curves. This estimation procedure for the risk factors gives the initial condition $v_k(t)$. Then, the postulated dynamics for the risk factors induce a dynamic for $G$, and hence of the forward variance.

Notice that in this approach, there is no relation with the underlying and its dynamic. For this reason, the possible processes are weakly constrained, and the parameters need to be estimated independently (say for example the characteristic times $\tau_k$). Another drawback of this approach is to rely on the empirical forward variance curve, and therefore a liquid option market is a prerequisite. 

Our choice of notations makes clear the formal analogy of the market model with the forecasts produced by a multi-component ARCH process. Except for the detailled shapes of the functions $w_k(\DT)$, the equations \ref{eq:ARCHForecast} and \ref{eq:twoCompStochVol} have the same structure. They are however quite different in their spirits as the $v_k$ are computed from the underlying time series in the ARCH approach, whereas in a market model approach the $v_k$  are estimated from the forward variance curve obtained from the option market. 
In other word, ARCH leads to a genuine forecast based on the underlying, whereas market model provides for a constraint fit of the empirical forward curve. Beyond this formal analogy, the dynamic for the risk factors are quite different as the ARCH approach leads to the unusual eq.~\ref{eq:dv_k} whereas market models use the familiar generic Gaussian process in eq.~\ref{eq:dv_k_general}.

\section{Comparison of the empirical implied, forecasted and realized volatilities}
\label{sec:empiricalComparison}
As explained in sec.~\ref{sec:forwardVarianceAndARCH}, a multi-components ARCH process provides us with a forecast for the realized volatility, and the forecast is directly related to the underlying process and its properties. At a given time $t$, there is three volatilities (implied, forecasted and realized) for each forecast horizon $\DT$. Essentially, the implied and forecasted volatilities are forecasts for the realized volatility. In this section, we investigate the relationship between these three volatilities and the forecast horizon $\DT$. 
When analyzing the empirical statistics and comparing these three volatilities, several factors should be kept in mind.
\begin{enumerate}

\item For short forecast horizons ($\DT = $ up to 10 days), the number of returns in $\DT$ is small and therefore the realized volatility estimator (computed with daily data) has a large variance.

\item The forecastability decreases with increasing $\DT$.

\item The forecast and implied volatilities are ``computed'' using the same information set, namely the history up to $t$. This is different from the realized volatility, computed using the information in the interval $[t, t+\DT]$. Therefore, we expect the distance between the forecast and implied to be the smallest.

At a more detailed level, the information set for the implied volatility is richer, because traders use intra-day information which helps building better forecasts, particularly for short risk horizons. 
This contrasts with all the present ARCH forecasts that are computed using only daily close prices.
From this difference on their actual information sets, the implied volatility can be expected to provide for a better forecast of the realized volatility.
 
\item The implied volatility has some particular idiosyncracies related to the option market, for example supply and demand, or the liquidity of the underlying necessary to implement the replication strategy. Similarly, an option bears a volatility risk, and a related volatility risk premium can be expected. These particular effects could bias the implied volatility upward.

\item From the raw options and underlying prices, the computations leading to the implied volatility are complex, and therefore error prone. 
This data quality problem is inherent to the original data provider and the option market, and is a reflect of the difficulty to compute clean and reliable implied volatility surfaces.
For stocks, the problem is made more difficult because of the dividents, the corporate events and the smaller liquidity.
For this reason, we present only the figures corresponding to two of the most liquid option markets.
The results have been checked with other FX rates, stock indexes and stocks, and are essentially valid for all underlyings.

\item The options are traded for fixed maturity time, whereas the convenient volatility surface is given for constant time to maturity. Therefore, some interpolation and extrapolation need to be done. 
As exchanged traded options are defined with one maturity per month, it is difficult to get reliable implied volatility for time to maturity smaller than one month. 

\item The ARCH based forecasts are dependent on the choice of the process and the associated parameters. 

\item As the forecast horizon increases, the dynamic of the volatility get slower and the actual number of independent volatility points decreases (as $1/\DT$). Therefore, the statistical uncertainty on the statistics are increasing with $\DT$.

\end{enumerate}

Because of the above points, each volatility has some peculiarities, and therefore we do not have a firm anchor point to base our comparison.  Given that we are on a floating ground, our goals are fairly modest. 
Essentially, we want to show that a process with one time scale is not good enough, and that the long-memory process provides for a good forecast with an accuracy comparable to the implied volatility. 
The processes used in the analysis are I-GARCH(1), I-GARCH(2) with two set of parameters and LM-ARCH. The equations for the processes are given in sec.~\ref{sec:ARCH}, with the values for the parameters.

\begin{figure}
\begin{center}
 \makebox[0.32\textwidth][c]{1.1.2002}\hspace{1ex}\makebox[0.32\textwidth][c]{1.1.2003}\hspace{1ex}\makebox[0.32\textwidth][c]{1.1.2004}\\
 \includegraphics[clip, width = 0.32\textwidth]{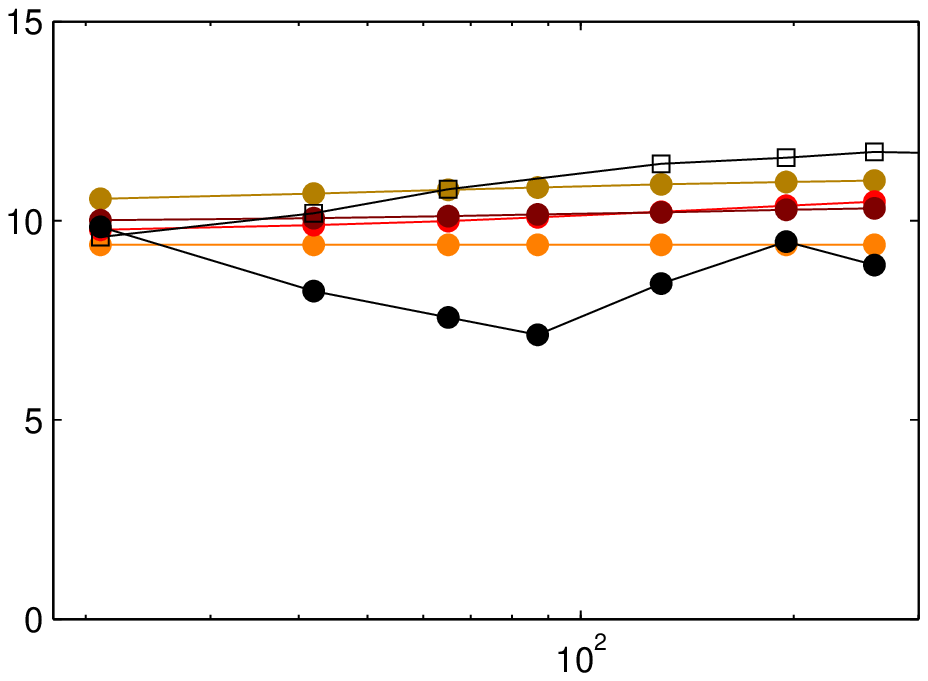}\hspace{1ex}
 \includegraphics[clip, width = 0.32\textwidth]{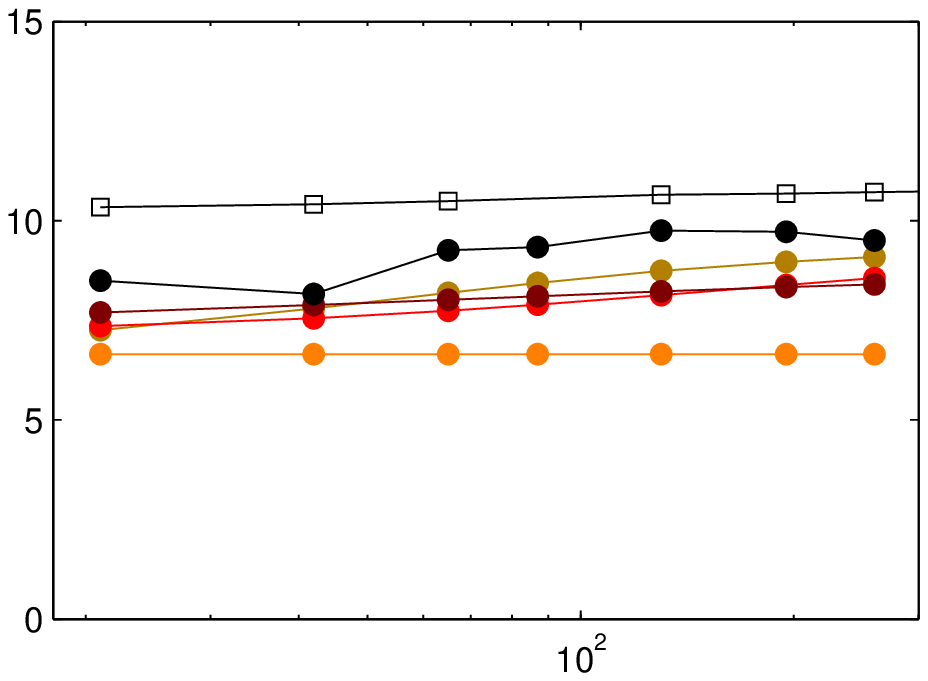}\hspace{1ex}
 \includegraphics[clip, width = 0.32\textwidth]{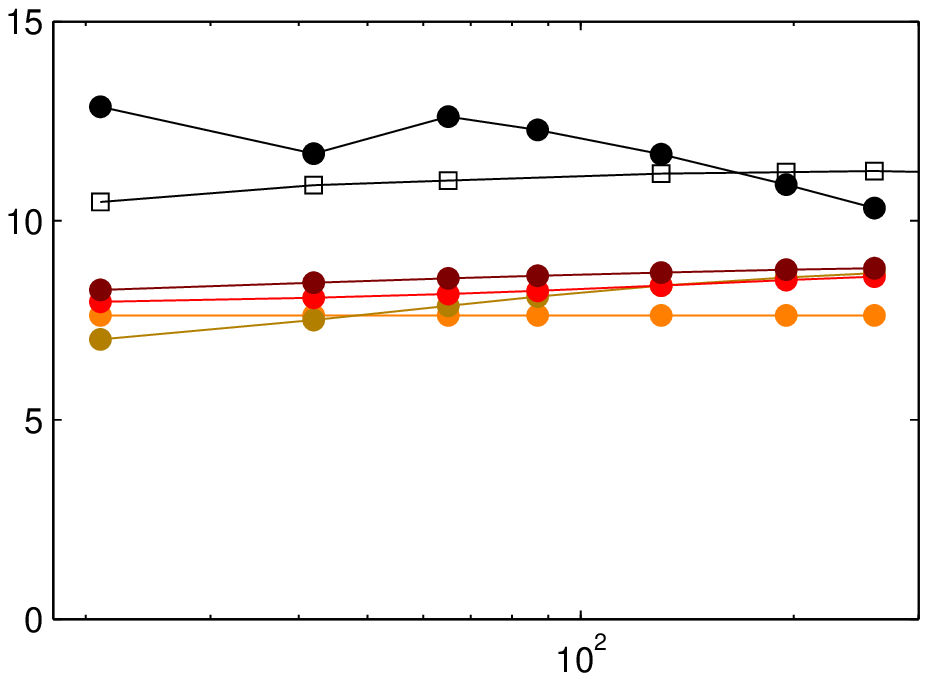}\\[2ex]
 \makebox[0.32\textwidth][c]{3.1.2005}\hspace{1ex}\makebox[0.32\textwidth][c]{3.1.2006}\hspace{1ex}\makebox[0.32\textwidth][c]{1.1.2007}\\
 \includegraphics[clip, width = 0.32\textwidth]{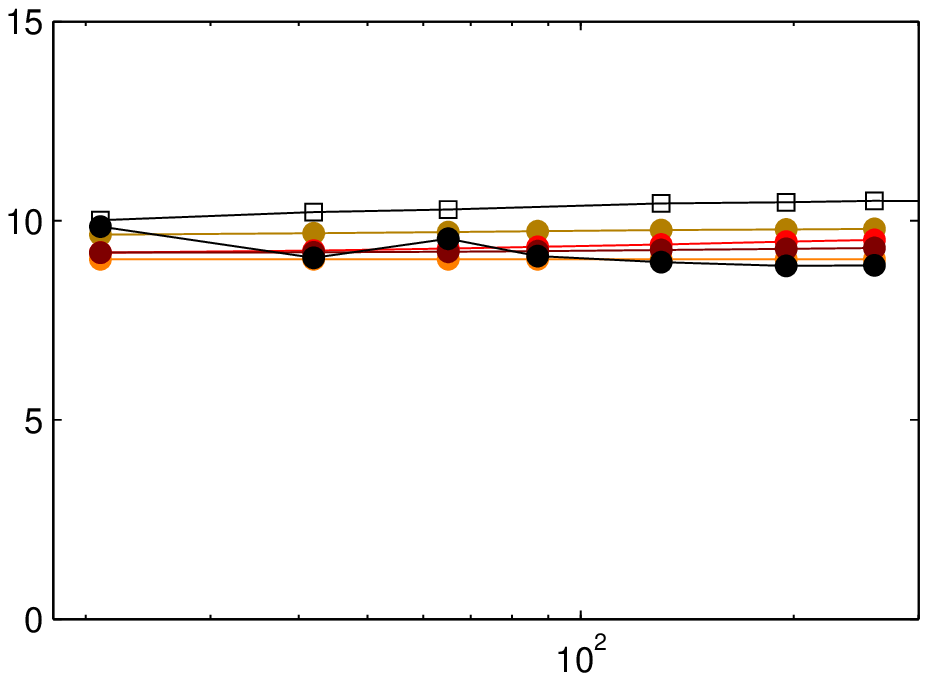}\hspace{1ex}
 \includegraphics[clip, width = 0.32\textwidth]{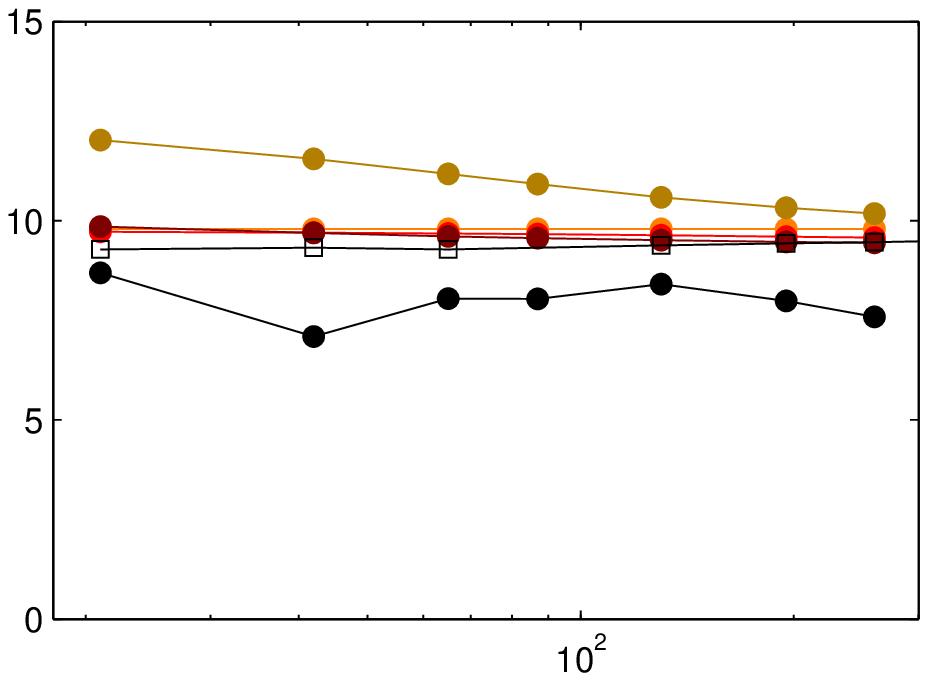}\hspace{1ex}
 \includegraphics[clip, width = 0.32\textwidth]{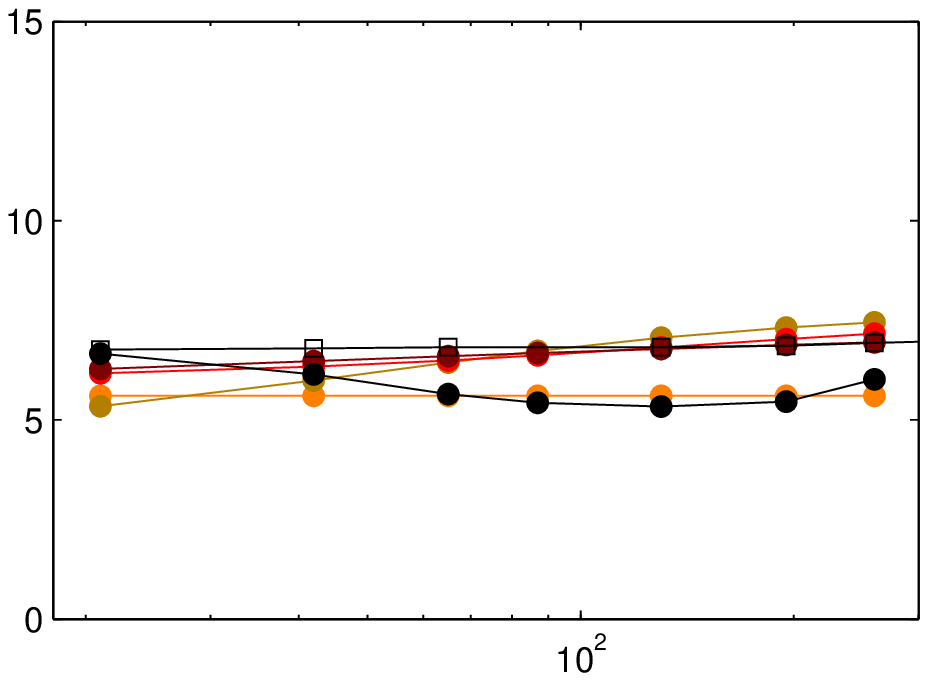}
 \end{center}
 \caption{The volatilities at the beginning of the years 2002 to 2007, for EUR/USD.
	The black curve with square symbols is the realized volatility, the black curve with full circle symbols is the implied volatility, and the color curve with full circle symbols is the forecast according to the various ARCH processes (with the same colors as below). 
	The vertical axis gives the annualized volatility in \%, the horizontal axis the forecast time interval $\DT$ in day.
}
  \label{fig:volatilitySnapshot}
\end{figure}
The best way to visualize the dynamic of the three volatilities would be to use a movie of the $\sigma[\DT]$ time evolution. 
Unfortunately, the present analogic paper does not allow for such medium, and we present instead 6 snapshots for EUR/USD in Figure~\ref{fig:volatilitySnapshot}.
Overall, the realized volatility has a weak term structure, although the global level changes significantly with time.
The implied volatility has more structures as function of the time to maturity, but this seems not always appropriate.
The term structures for the ARCH forecasts are in line with the implied volatility, with essentially a weak term structure.
The I-GARCH(1) process has a constant term structure, and this explains why its forecasting performances are indeed very good compared to more complex processes.
Beyond a qualitative assessement of the term structure, the various forecasts for the realized volatility are difficult to rank, but clearly the ARCH forecasts are close to the target and compare well with the implied volatility. 

\begin{figure}
 \begin{center}
 \includegraphics[clip, width = 0.98\textwidth]{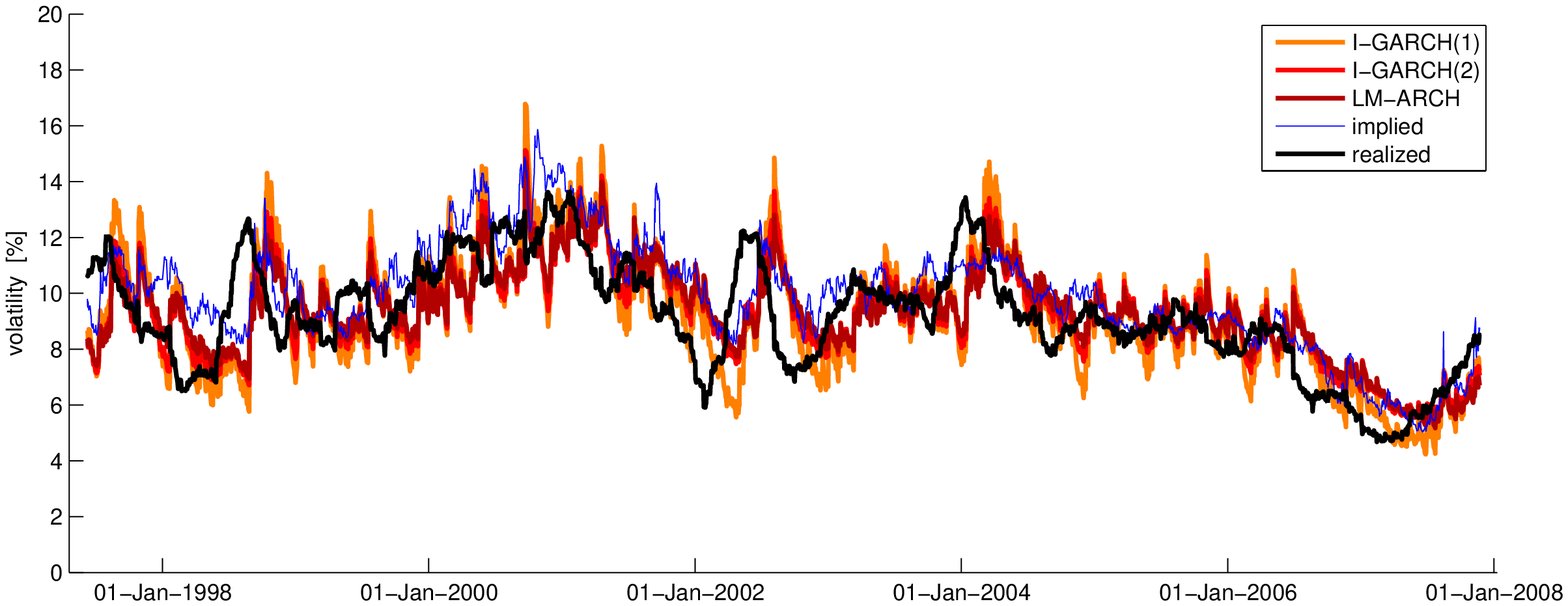}\\[1ex]
 \includegraphics[clip, width = 0.98\textwidth]{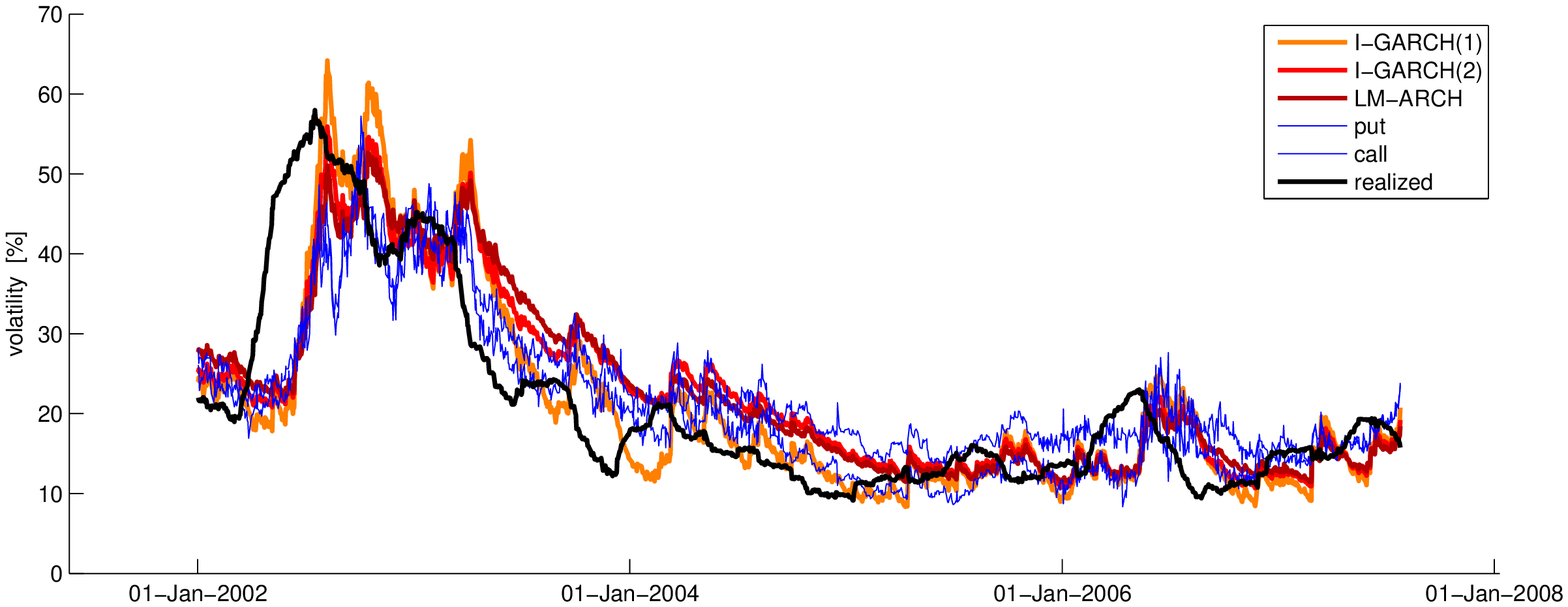}
 \end{center}
  \caption{The volatilities time series for the USD/EUR (upper panel) and DAX (lower panel), for a 3 months forecast horizon.
	For the DAX data, the implied volatility is given for the put and call options (blue curves).}
  \label{fig:timeSeries}
\end{figure}
The statistics are presented for two time series, the USD/EUR foreign exchange rate and the DAX stock index.
The time series for the volatilities are shown on fig.~\ref{fig:timeSeries} for a 3 months forecast horizon. The time series are not very long ($\sim$10 years for USD/EUR, $\sim$6 years for DAX). This clearly makes statistical inferences difficult, as the effective sample size is fairly small. 
The lagging behavior of the forecast and implied volatility with respect to the realized volatility is clearly observed.
For the DAX, the data sample contains an abrupt drop in the realized volatility at the beginning of 2003. 
This pattern was difficult to capture for the models with long term mean reversion.

For the statistics, all the horizontal and vertical scales are identical, and the colors are fixed for a given process. 
The graphs are presented for the mean absolute error (MAE)
\begin{equation}
  \text{MAE}(x, y) = \frac{1}{n} \sum_t \left| x(t) - y(t) \right|
\end{equation}
where $n$ is the number of term in the sum. 
Other measures of distance like root mean square error, or the MAE for $\ln(\sigma)$, give very similar figures. 

\begin{figure}
\begin{center}
 \makebox[0.47\textwidth][c]{\bf I-GARCH(1)}\hspace{2ex}\makebox[0.47\textwidth][c]{\bf I-GARCH(2) ~~~parameter set 1}\\
 \includegraphics[clip, width = 0.47\textwidth]{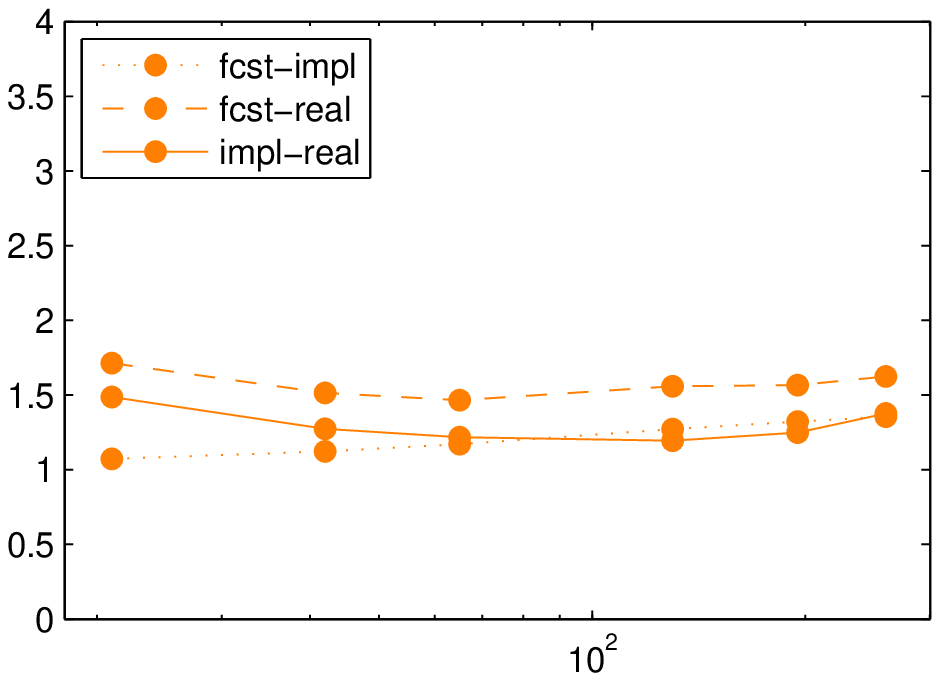}\hspace{2ex}
 \includegraphics[clip, width = 0.47\textwidth]{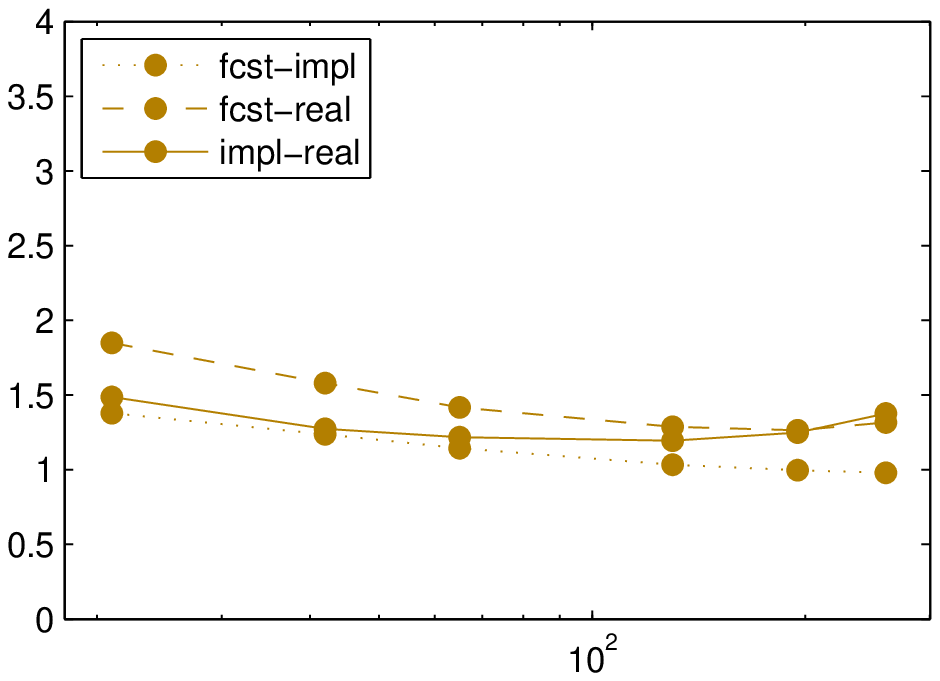}\\[2ex]
 \makebox[0.47\textwidth][c]{\bf I-GARCH(2)~~~parameter set 2}\hspace{2ex}\makebox[0.47\textwidth][c]{\bf LM-ARCH}\\
 \includegraphics[clip, width = 0.47\textwidth]{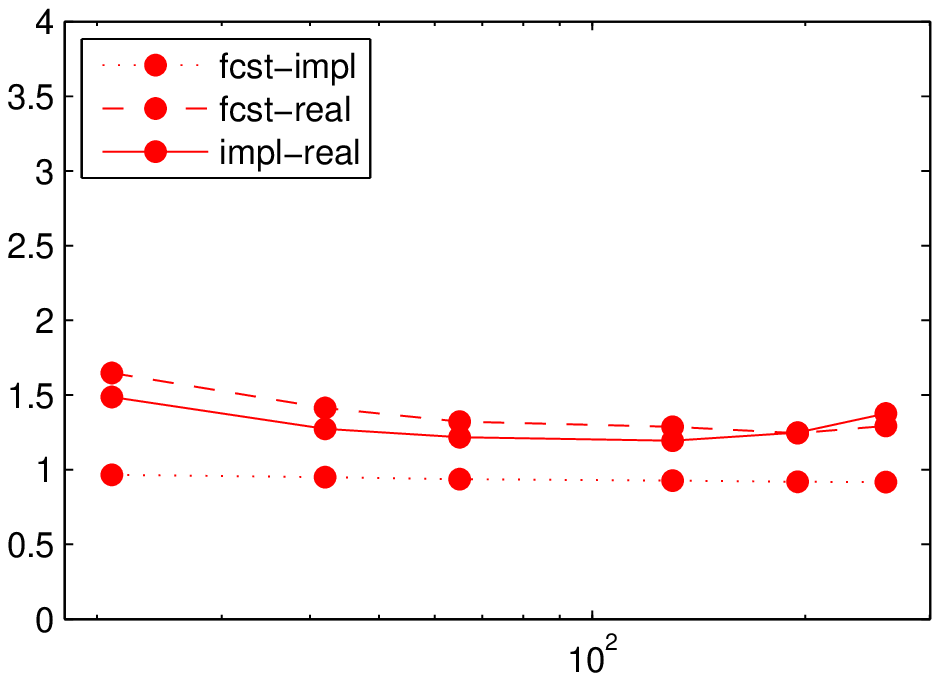}\hspace{2ex}
 \includegraphics[clip, width = 0.47\textwidth]{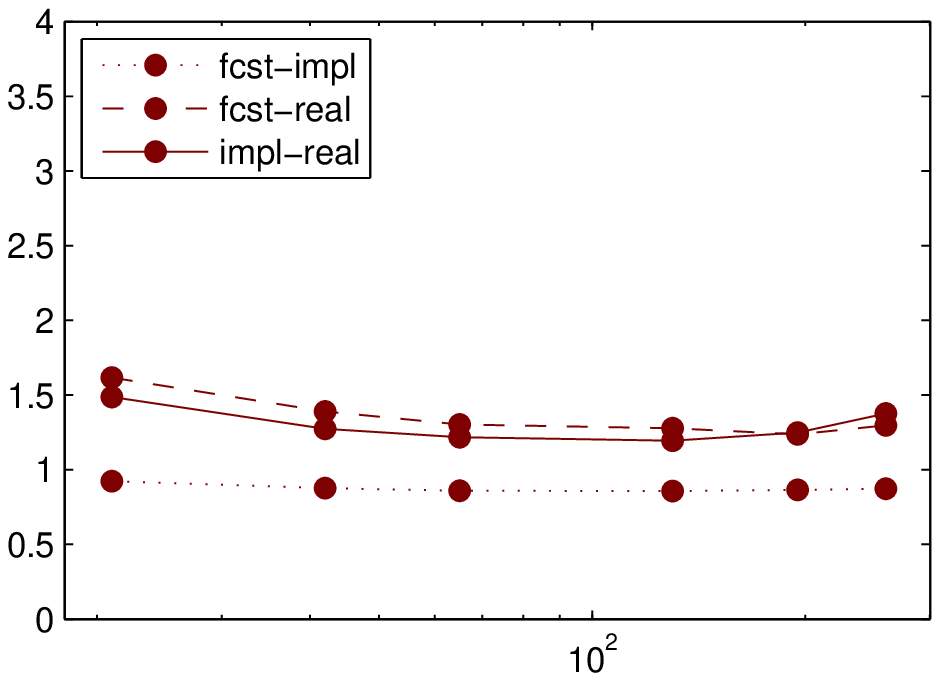}
 \end{center}
 \caption{The MAE distances between volatility pairs for different forecasts: I-GARCH(1) (upper left, red), I-GARCH(2) parameters 1 (upper right, blue), I-GARCH(2) parameters 2 (lower left, blue) and LM-ARCH (lower right, black). 
The vertical axis gives the MAE for the annualized volatility in \%, the horizontal axis the forecast time interval $\DT$ in day.
The data is EUR/USD.}
  \label{fig:forecatsModels}
\end{figure}
The overall relationship betwen the three volatilities can be understood on figure~\ref{fig:forecatsModels}.
The pair of volatilities with the closest relationship is the implied and forecasted volatilities, because they are build upon the same information set. 
The distance with the realized volatility is larger, with similar values for implied-realized and forecast-realized. This shows that it is quite difficult to assert which one of the implied and forecasted volatility provides for a better forecast of the realized volatility. 
All the distances have a global U-shape form as function of $\DT$.
This originates in the points 1 and 2 above, which leads to a minimum between 2 to 6 months for the distances. 
The distance is larger for shorter $\DT$ because of the bad estimator for the realized volatility, and larger for longer $\DT$ because of the decreasing forecastability. 
The time structures of the ARCH processes impact the distances between the forecasted and implied volatility (dotted line),
and the relation between process structure and forecast quality discussed in the next paragraph.

\begin{figure}
\begin{center}
 \makebox[0.47\textwidth][c]{\bf EUR/USD ~~~~forecast-implied}\hspace{2ex}\makebox[0.47\textwidth][c]{\bf EUR/USD ~~~~forecast-realized}\\
 \includegraphics[clip, width = 0.47\textwidth]{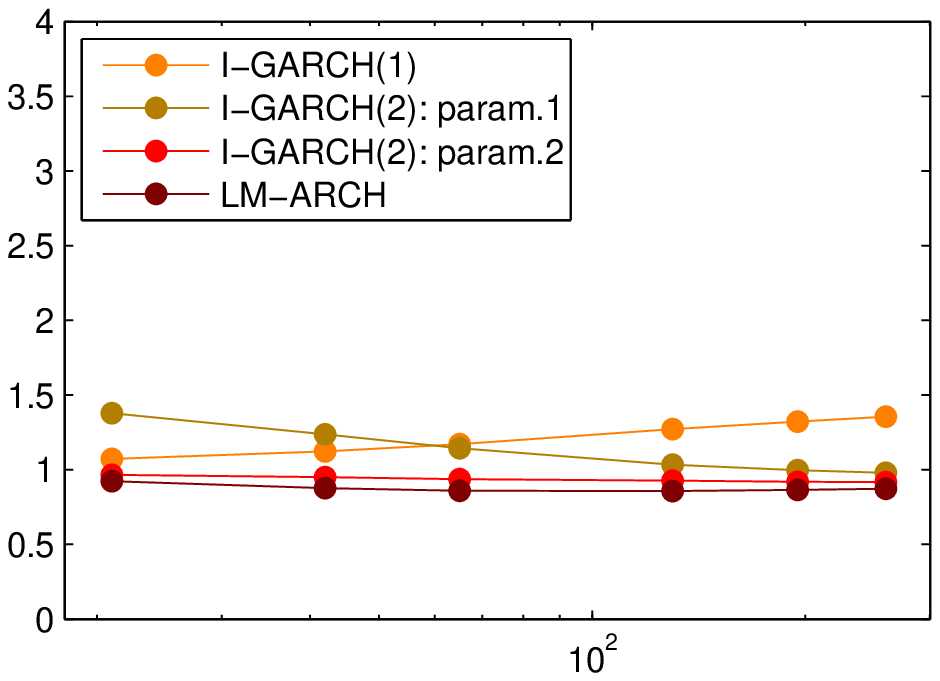}\hspace{2ex}
 \includegraphics[clip, width = 0.47\textwidth]{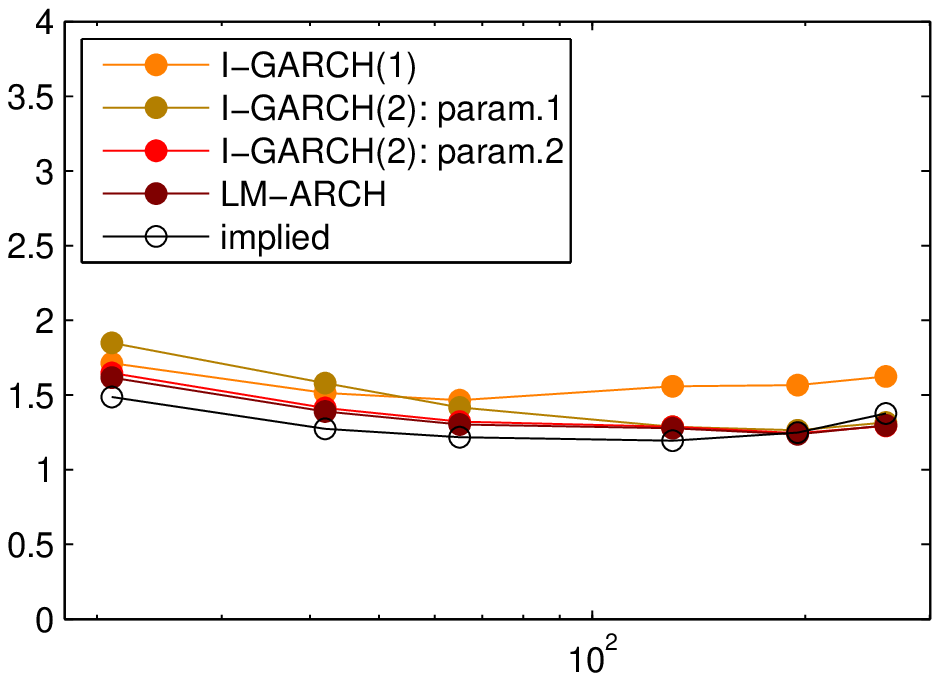}\\[2ex]
 \makebox[0.47\textwidth][c]{\bf DAX ~~~~forecast-implied}\hspace{2ex}\makebox[0.47\textwidth][c]{\bf DAX ~~~~forecast-realized}\\
 \includegraphics[clip, width = 0.47\textwidth]{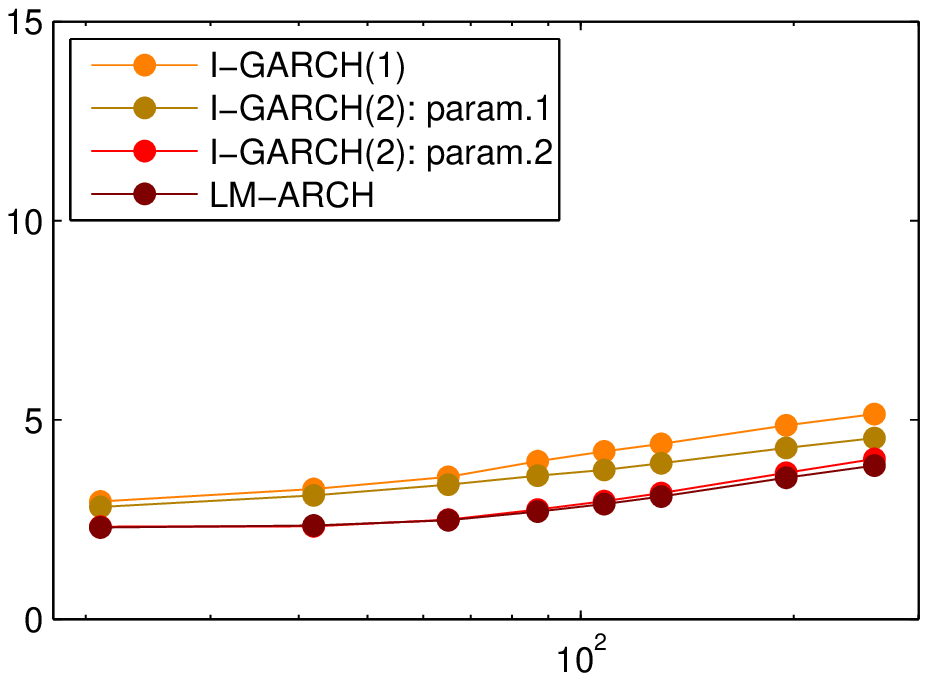}\hspace{2ex}
 \includegraphics[clip, width = 0.47\textwidth]{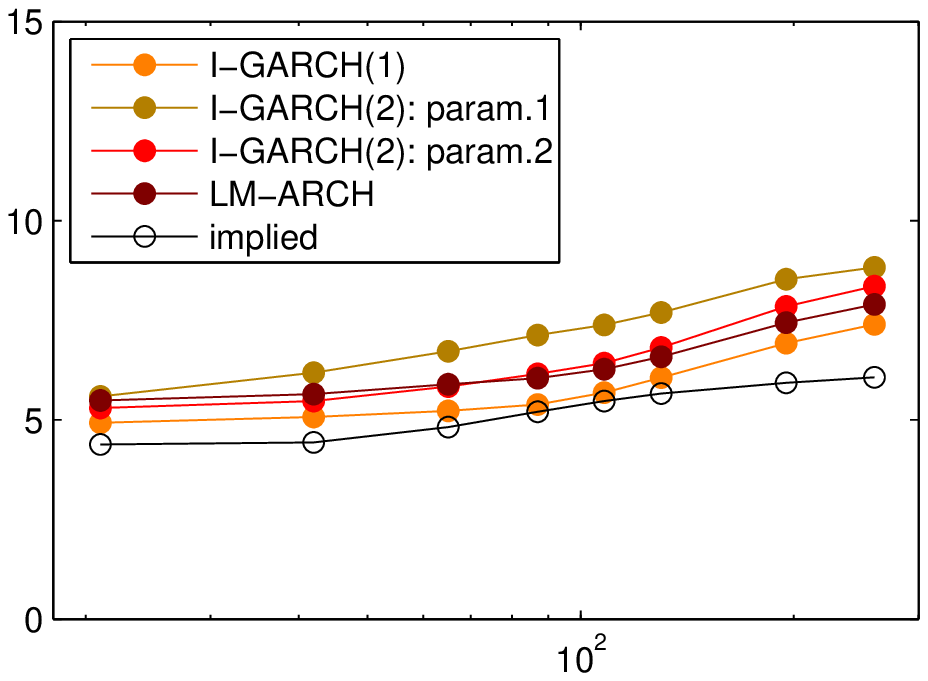}
\end{center}
\caption{The MAE distances between volatility pairs: forecast-implied (left) and forecast-realized (right).
     The upper figures are for EUR/USD, the lower figure for the DAX stock index.
The vertical axis gives the MAE for the annualized volatility in \%, the horizontal axis the forecast time interval $\DT$ in day.}
  \label{fig:dist_vol}
\end{figure}
The figure~\ref{fig:dist_vol} shows the distances for given volatility pairs, depending on the process used to build the forecast. 
The forecast-implied distance shows clear difference between processes (left panels).
The I-GARCH(1) process is lacking mean reversion, an important feature of the volatility dynamic. 
The I-GARCH(2) process with parameter set 1 is handicapped by the too short characteristic time for the first EMA (4 days); this leads to a noisy volatility estimator and subsequently to a noisy forecast.
The same process with a longer characteristic time for the first EMA (16 days,  parameter set 2) shows much improved performance up to a time horizon comparable to the long EMA (512 days). Finally, the LM-ARCH produces the best forecast. 
As the forecast becomes better (1 time scale $\rightarrow$ 2 time scales $\rightarrow$ multiple time scales), the distance between the implied and forcasted volatilities decreases. 
For EUR/USD, the mean volatility is around 10\% (the precise value depending on the volatility and time horizon), and the MAE is in the 1 to 2\% range. 
This shows that in this time to maturity range, we can build a good estimator of the ATM implied volatility based only on the underlying time series.

The distance forecast-realized is larger than the forecast-implied volatility (right panel), with the long memory process giving the smallest distance. The only exception is the I-GARCH(1) process applied to the DAX time series, due to the particular abrupt drop in the realized volatility at early 2003. 
This shows the limit of our analysis due to the fairly small data sample, and longer time series for implied volatility are required to gain more statistical power.
Given the limited sample size, a cross sectional study over 9 other time series shows consistent results.

\section{ Conclusion }
The ``m\'{e}nage \`{a} 3'' between the forecasted, implied and realized volatilities is quite a complex affair, where each participants have their own character. 
The salient outcome is that the forecasted and impled volatilities have the closest relationship,  while the realized volatility is more distant as it incorporates a larger information set. 
This picture is dependent to some extend on the quality of the volatility forecast: the multi-scale dynamic of the long memory ARCH process is seen to capture correctly the dynamic of the volatility, while the I-GARCH(1) process is not rich enough in its time scale structures. 
This conclusion falls in line with the risk methodology developed in \cite{Zumbach.RM2006_fullReport}, where the same long memory process is shown to capture correctly the lagged correlation for the volatility. 

The connection with the market model for the forward variance shows the parallel in the structure of the volatility forecasts provided by both approaches. 
However, their dynamics are very different (postulated for the forward volatility market models, induced by the ARCH structure for the multi-components ARCH processes). 
Moreover, the volatility process induced by the ARCH equations is of a different type than the usual price process, because the random term is of order $\dt$ instead of $\sqrt{\dt}$ used in diffusive equations. 
This emphasize a fundamental difference between price and volatility processes. 
 A clear advantage of the ARCH approach is to deliver a forecast based only on the properties of the underlying time series, with a minimal number of parameters that need to be estimated (none in our case as all the parameters correspond to the values used in \cite{Zumbach.RM2006_fullReport}). 
This point brings us to a nice and simple common framework to evaluate risks as well as a good approximation for the implied volatilities of at-the-money options. 

The natural extension of this work is to study the whole implied volatility surface.
As the backbone is essentially under control, the perpendicular direction needs to be studied, namely the volatility smile should be related to the underlying behavior. 
Due to the heteroscedasticity, any multi-component ARCH process will capture some (symmetric) smile. 
Moreover, fat tail innovations will make the smile stronger, as the process becomes increasingly distant from a Gaussian random walk. 
Yet, adding an asymmetry in the smile, as observed for stocks and stock indexes, requires to enlarge the family of process to capture asymmetries in the distribution of returns. 
This is left for further work.


\newpage
\bibliography{backbone}

\begin{thebibliography}{}

\bibitem[Bergomi, 2005]{Bergomi}
Bergomi, L. (2005).
\newblock Smile dynamics ii.
\newblock {\em Risk}, 18:67--73.

\bibitem[Buehler, 2006]{Buehler}
Buehler, H. (2006).
\newblock Consistent variance curve models.
\newblock {\em Finance and Stochastics}, 10:178--203.

\bibitem[Dacorogna et~al., 1998]{Dacorogna.1998}
Dacorogna, M.~M., M\"{u}ller, U.~A., Olsen, R.~B., and Pictet, O.~V. (1998).
\newblock Modelling short-term volatility with {GARCH} and {HARCH} models.
\newblock {\em published in ``Nonlinear Modelling of High Frequency Financial
  Time Series'' edited by Christian Dunis and Bin Zhou, John Wiley,
  Chichester}, pages 161--176.

\bibitem[Engle and Bollerslev, 1986]{EngleBollerslev.1986}
Engle, R.~F. and Bollerslev, T. (1986).
\newblock Modelling the persistence of conditional variances.
\newblock {\em Econometric Reviews}, 5:1--50.

\bibitem[Gatheral, 2007]{Gatheral}
Gatheral, J. (2007).
\newblock Developments in volatility derivatives pricing.
\newblock Presentation at ``Global derivative'', Paris, May 23.

\bibitem[Lynch and Zumbach, 2003]{LynchZumbach}
Lynch, P. and Zumbach, G. (2003).
\newblock Market heterogeneities and the causal structure of volatility.
\newblock {\em Quantitative Finance}, 3:320--331.

\bibitem[Nelson, 1990]{Nelson.1990}
Nelson, D. (1990).
\newblock Arch model as diffusion approximation.
\newblock {\em Journal of Econometrics}, 45:7--38.

\bibitem[Poon, 2005]{Poon.2003}
Poon, S.-H. (2005).
\newblock {\em Forecasting financial market volatility}.
\newblock Wiley Finance.

\bibitem[Zumbach, 2004]{Zumbach.LongMemory}
Zumbach, G. (2004).
\newblock Volatility processes and volatility forecast with long memory.
\newblock {\em Quantitative Finance}, 4:70--86.

\bibitem[Zumbach, 2006]{Zumbach.RM2006_fullReport}
Zumbach, G. (2006).
\newblock The riskmetrics 2006 methodology.
\newblock Technical report, RiskMetrics Group.
\newblock Available at www.riskmetrics.com.

\bibitem[Zumbach and Lynch, 2001]{ZumbachLynch}
Zumbach, G. and Lynch, P. (2001).
\newblock Heterogeneous volatility cascade in financial markets.
\newblock {\em Physica A}, 298(3-4):521--529.

\end{thebibliography}
\bibliographystyle{apalike}

\end{document}